# Statistical Modeling of Pipeline Delay and Design of Pipeline under Process Variation to Enhance Yield in sub-100nm Technologies*

Animesh Datta, Swarup Bhunia, Saibal Mukhopadhyay, Nilanjan Banerjee, and Kaushik Roy
Dept. of ECE, Purdue University, West Lafayette, IN, 47907, USA
<adatta, bhunias, sm, nbanerje, kaushik> @ecn.purdue.edu

## Abstract

*Operating frequency of a pipelined circuit is determined by the delay of the slowest pipeline stage. However, under statistical delay variation in sub-100nm technology regime, the slowest stage is not readily identifiable and the estimation of the pipeline yield with respect to a target delay is a challenging problem. We have proposed analytical models to estimate yield for a pipelined design based on delay distributions of individual pipe stages. Using the proposed models, we have shown that change in logic depth and imbalance between the stage delays can improve the yield of a pipeline. A statistical methodology has been developed to optimally design a pipeline circuit for enhancing yield. Optimization results show that, proper imbalance among the stage delays in a pipeline improves design yield by 9% for the same area and performance* (*and area reduction by about 8.4% under a yield constraint*) *over a balanced design.*

## 1. Introduction

Increasing inter-die and intra-die variations in the process parameters, such as channel length, width, threshold voltage etc., result in large variation in the delay of logic circuits [1]. Consequently, estimating circuit performance and designing high-performance circuits with high yield (probability that the design will meet certain delay target) under parameter variations have emerged as serious design challenges in sub-100nm regime [1, 2, 5]. Statistical analysis of delay and techniques to enhance yield in combinational circuits have been proposed [2, 3]. In the high-performance design, the throughput is primarily improved by pipelining the data and control paths [4]. In a synchronous pipelined circuit, the throughput is limited by the slowest pipe segment (segment with maximum delay) [4]. Under parameter variations, as the delays of all the stages vary considerably, the slowest stage is not readily identifiable. The variation in the stage delays thus result in variation in the overall pipeline delay (which determines the clock frequency and throughput). Thus, a statistical delay model is necessary to predict the delay distribution of the pipeline.

Traditionally, the pipeline clock frequency has been enhanced by: a) increasing the number of pipeline stages, which, in essence, reduces the logic depth and hence, the delay of each stage [4]; and b) balancing the delay of the pipe stages, so that the maximum of stage delays are optimized [4]. However, it has been shown that if intra-die parameter variation is considered, reducing the logic depth increases the variability (defined as the ratio of standard deviation and mean) [5]. The effect of balancing the stage delays on the overall delay under parameter variation also needs to be analyzed. Thus, traditional deterministic approaches for maximizing the pipeline throughput need to be reinvestigated to understand their effect on pipeline yield under parameter variations.

In this paper, we have developed statistical delay models and proposed a statistical design flow for enhancing the yield of a pipelined design considering inter-die and intra-die variations. In particular, in this paper, we have developed:

- Analytical models to estimate the mean and the standard deviation of the overall pipeline delay from the individual stage delay distributions.
- Analytical models to predict yield of a pipelined design.
- Analysis of the effect of logic depth and imbalance in the stage delays and correlation among the different stage delays on the yield of a pipelined design.
- A statistical design methodology to minimize the area (or power) of a pipelined design under a target yield and performance constraint.

Our analysis shows that, under large intra-die process variations reducing the logic depth decreases the yield, which is in accordance with [5]. However, if inter-die process variation is dominant, the traditional approach of reducing the logic depth, results in better yield. It has also been shown that, balancing delays of different pipeline stages *does not necessarily maximize* the yield. Under parameter variations, proper imbalance among the stage delays result in yield improvement. We have come up with a simple heuristic (based on the area vs. delay trend of a circuit) to systematically incorporate proper imbalance among pipeline stages, so that yield can be improved with minimum area penalty. The heuristic is used to improve yield of an example 4 stage pipelined design by 9% from a *balanced design* (with stages *independently optimized* for equal stage delay). Based on the above observation, we have designed a fast gate-level sizing algorithm for the *complete pipeline* to minimize total area under a yield constraint. This optimization step reduces the area by 8.4% (while ensuring a 80% target yield) compared to a balanced design.

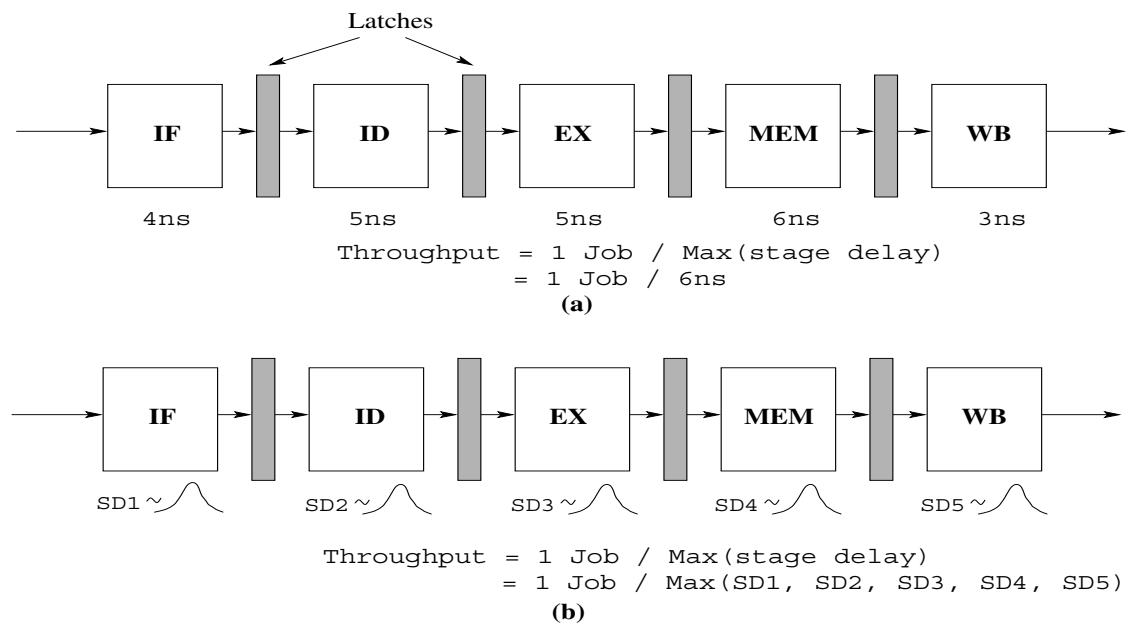

**Figure 1: A simple 5-stage pipeline and its throughput for (a) static delay model, (b) statistical delay model**

* The work is sponsored in part by Marco Gigascale Systems Research Center (GSRC), Intel Corp. and Semiconductor Research Corp. (SRC).

## 2. Yield estimation of a pipeline design under parameter variation

The delay of a pipeline is determined by the slowest pipe stage (Fig. 1(a)) [4]. The variation in process parameters results in a significant variation in the delay of each pipe stage and hence, the overall pipeline delay [1, 5].

### 2.1. Variation in pipeline delay and yield

The delay of a pipeline stage ($SD$) consists of the clock-to-Q delay of the latch ($T_{C\text{-}Q}$), propagation delay through the combinational logic ($T_{comb}$) and the setup time ($T_{setup}$) [4]. Both the inter-die and the intra-die distribution in process parameters result in the variation of the delay of the pipe stages. Inter-die distribution shifts the delay of each stage in the same direction (i.e. either all increases or all decreases). Due to intra-die distribution, the delay of different stages may shift in different directions. The random component of the intra-die distribution (e.g. $V_{th}$ variation due to Random Dopant Fluctuation [6]) makes the delays of the different stages in a pipeline completely *independent* of each other (uncorrelated stage delays). On the other hand, the systematic variation in the parameters (e.g. spatially correlated $W$, $L$, $T_{ox}$ variations) makes the delays of the different stages correlated [1]. Moreover, the stages can also be electrically correlated if the delay of the combinational logic of one stage affects the delay of the subsequent stage [6].

Let us consider a pipeline consisting of N stages (Fig. 1). If $SD_i$ denotes the delay of $i$-th stage, then the overall pipeline delay ($T_P$) is the maximum of N individual stage delays and is given by [4]:

$$T_P = \underset{i=1,\dots,N}{Max}(SD_i) = \underset{i=1,\dots,N}{Max}(T_{C-Q^i} + T_{Comb^i} + T_{Setup^{i+1}}) \quad (1)$$

The stage delays can be represented as *correlated Gaussian random* variables ($SD_i \sim N(\mu_i, \sigma_i)$) with mean $\mu_i$ and standard deviation $\sigma_i$ [1, 3]. If we neglect the spatial and the electrical correlation, the stage delays can be considered as independent random variables. The overall pipeline delay $T_P$ given in (1) will also be a random variable with the mean $\mu_T$ and the standard deviation $\sigma_T$.

The probability ($P_D$) that the pipelined design will meet a specific delay requirement (say $T_{TARGET}$) is given by:

$$P_D = Yield = \Pr\{T_P < T_{TARGET}\} = \Pr\{\max_{i=1,\dots,N} SD_i < T_{TARGET}\} \quad (2)$$

We define this probability as a measure of the yield of the pipelined design. In this section we will discuss methods to analytically estimate the overall delay distribution of the pipeline (i.e. $\mu_T$, $\sigma_T$) and the yield (i.e. $P_D$).

### 2.2. Estimation of distribution of pipeline delay

The mean ($\mu_T$) and the standard deviation ($\sigma_T$) of the pipeline delay ($T_P$) depend on the mean ($\mu_i$) and the standard deviation ($\sigma_i$) of the each stage. A lower bound on $\mu_T$ is given by (using the Jensen's inequality [7, 8]):

$$\mu_T = E[T_P] = E[\max_{i=1,\dots,N} SD_i] \geq \max_{i=1,\dots,N}\{E[SD_i]\} \quad (3)$$

The above equation states that the mean of the overall pipeline delay will be larger than the maximum of the mean delay of all the stages. To obtain an exact estimate of $\mu_T$ and $\sigma_T$ we approximate $T_P$ as [8]:

$$\begin{aligned} T_P &= \max(SD_1, SD_2, \dots, SD_{n-1}, SD_n) \\ &= \max(SD_1, SD_2, \dots, \max(SD_{n-1}, SD_n)) \\ &= \max(SD_1, SD_2, \dots, \max(SD_{n-2}, N_{n-1,n})) \end{aligned} \quad (4)$$

where, $N_{n-1,n}$ represents the normal approximation to $max$ ($SD_{n-1}$, $SD_n$). The mean ($\mu_{n-1,n}$) and the standard deviation ($\sigma_{n-1,n}$) of $N_{n-1,n}$ can be approximated as [8]:

$$\begin{aligned} & m_1 = \mu_n \Phi(\alpha) + \mu_{n-1}\Phi(-\alpha) + a\varphi(\alpha) \\ & m_2 = (\mu_n^2 + \sigma_n^2)\Phi(\alpha) + (\mu_{n-1}^2 + \sigma_{n-1}^2)\Phi(-\alpha) \\ & \quad + (\mu_n + \mu_{n-1}) a\varphi(\alpha) \\ & where \;\; \alpha = (\mu_n - \mu_{n-1})/a \;\; ; \;\; a^2 = \sigma_n^2 + \sigma_{n-1}^2 - 2\sigma_n\sigma_{n-1}\rho_{n-1,n} \\ & Mean \; of \; N_{n-1,n} = \mu_{n-1,n} = m_1 \\ & Std.\,Deviation = \sigma_{n-1,n} = \sqrt{m_2 - m_1^2} \end{aligned} \quad (5)$$

where, $\Phi$ represents the Cumulative Distribution Function (CDF) and $\varphi$ represents the Probability Distribution Function (PDF) ($\varphi(\alpha) = (2\pi)^{-1/2}\exp(-\alpha^2/2)$) of a standard normal ($\mu = 0$ and $\sigma = 1$) Gaussian variable [8]. The correlation coefficient between $SD_{n-1}$ and $SD_n$ is given by $\rho_{n-1,n}$. The correlation coefficient between $N_{n,n-1}$ and $SD_{n-2}$ (say $\rho(SD_{n-2}, N_{n-1,n})$) is given by [8]:

$$\rho\left[SD_{n-2}, N_{n,n-1}\right] = \left[\sigma_n \rho_{n-2,n}\Phi(\alpha) + \sigma_{n-1}\rho_{n-2,n}\Phi(-\alpha)\right]/\sigma_{n-2} \quad (6)$$

$\rho(SD_{n-2}, N_{n-1,n})$ is used to estimate the mean and the standard deviation of max($SD_{n-2}$, $N_{n,n-1}$). The above process is repeated by taking two variables at a time and finally $\mu_T$ and $\sigma_T$ are estimated.

### 2.3. Estimation of yield

Using (2) the probability of delay failure ($P_D$) or yield can be estimated as [7]:

$$P_D = \Pr\{\max_{i=1,\dots,N} SD_i < T_{TARGET}\} = \Pr\{\bigcap_{i=1,\dots,N}(SD_i < T_{TARGET})\} \quad (7)$$

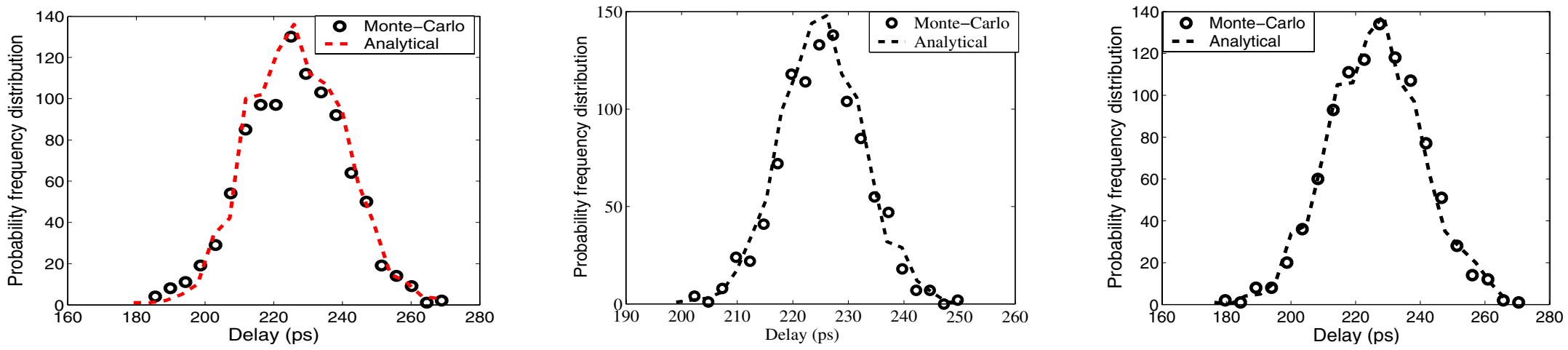

(a) (b) (c)

**Figure 2: Delay distribution of a 12-stage inverter chain pipeline (stage logic depth = 10) under process variation with (a) only random Intra-die variation, (b) only Inter-die variation, (c) Inter- and Intra-die variation with both random and systematic components**

The exact solution of (7) is possible by assuming the stage delays ($SD_i$) to be *independent* Gaussian random variables, as shown below [7]:

$$P_D = \Pr\{\bigcap_{i=1,\ldots,N}(SD_i < T_{TARGET})\} = \prod_{i=1}^{n}\Phi\left(\frac{T_{TARGET}-\mu_i}{\sigma_i}\right) \quad (8)$$

If the variables are correlated such a simplification is not possible. To estimate $P_D$ considering correlated $SD_i$s, we approximate the overall pipeline delay ($T_P$) as a Gaussian random variable (with $\mu_T$ and $\sigma_T$ estimated using (5)). Using this assumption $P_D$ is given by [7]:

$$P_D = \Pr\{T_D \le T_{TARGET}\} = \Phi\left(\frac{T_{TARGET}-\mu_T}{\sigma_T}\right) \quad (9)$$

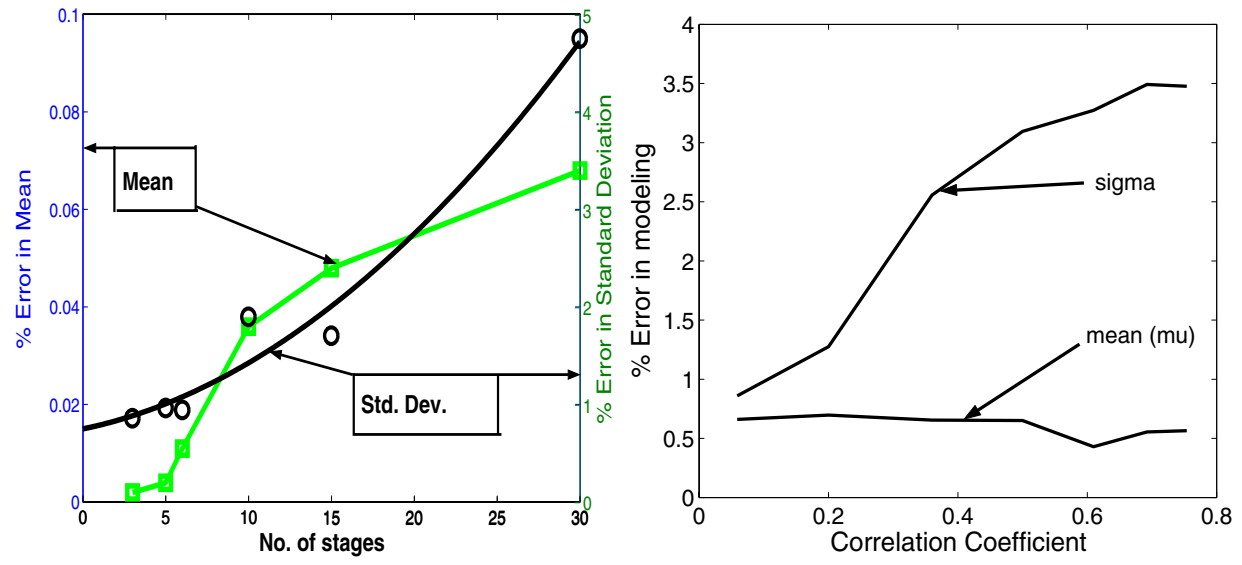

**Figure 3: Trend in modeling error with (a) number of stages and (b) correlation coefficient**

## 2.4. Model verification

We have estimated the delay distributions in different pipeline structures (number of pipe stage × logic depth in a stage). The models have been first verified by Monte-Carlo simulation of inverter chain pipelines with the transmission gate Master-slave flip-flops in 70nm BPTM [9]. SPICE Monte-Carlo simulation is first used to determine the mean and the standard deviation of the delay of each stage. The simulated $\mu_i$ and $\sigma_i$ values for each stage are then fed into the proposed model to determine the distribution of the pipeline delay. It can be observed that, the delay distribution predicted by the proposed model closely matches the SPICE simulation result of a 5x8 pipeline for: (a) only random intra-die variations (i.e. stage delays are considered as independent) (Fig. 2(a)) (b) only inter-die variations (i.e. stage delays are perfectly correlated) (Fig. 2(b)) and (c) both inter and intra-die distributions with spatial correlation (i.e. stage delays are partially correlated) (Fig. 2(c)). Close match have also been observed for several other pipeline configurations (Table-I). The major source of error in the proposed modeling method is the assumption that the maximum of two Gaussian variables is also a Gaussian one

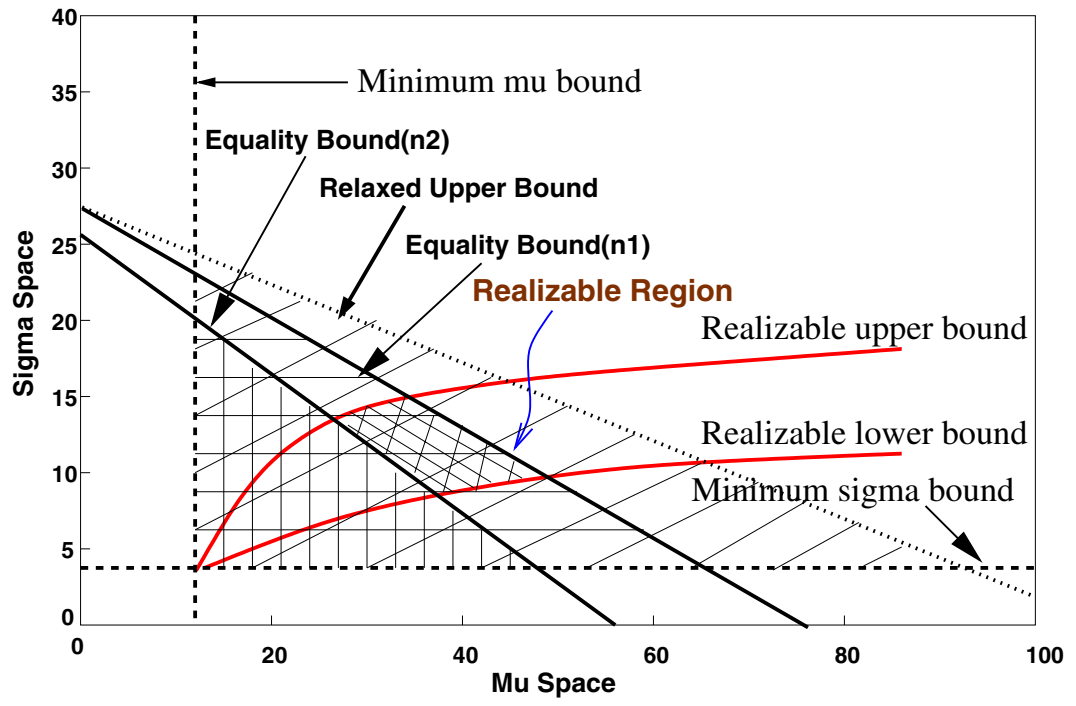

**Figure 4: Range of permissible mean and standard deviation for each stage to meet a target yield**

(i.e. $N_{n-1,n}$ is Gaussian) [7, 8]. This assumption is valid only for *two uncorrelated* variables [7, 8]. Hence, the errors in the estimation of the mean and the standard deviation are expected to increase with the increase in the number of pipeline stages (Fig. 3(a)) and the correlation between the stage delays (Fig. 3(b)) [7, 8]. It can be observed that the increase in error in the standard deviation is more significant in both cases. However, in all cases the error in the standard deviation and the mean is less than 3% and 0.2%, respectively. The modeling error also depends on the ordering of the variables $SD_i$ in (2). It has been shown that the error is the minimum if the variables are ordered in increasing (or decreasing) sequence of their means [7]. We have used this ordering in our estimation to minimize the modeling error.

**Table-I Modeling and simulation results of delay distribution and yield for different pipeline configurations**

| Pipeline Config. | Target Delay (ps) | Monte-Carlo | | | Analytical Model (ps) | | |
|---|---|---|---|---|---|---|---|
| | | $\mu_T$ (ps) | $\sigma_T$ (ps) | Yield (%) | $\mu_T$ (ps) | $\sigma_T$ (ps) | Yield (%) |
| 8 × 5 | 160 | 155 | 2.82 | 96 | 154 | 2.68 | 98.62 |
| 5 × 8 | 200 | 198 | 3.27 | 78 | 198 | 2.72 | 77.72 |
| 5 × * | 215 | 210 | 3.67 | 92 | 210 | 3.42 | 93.00 |
| 5 × 8 inter | 370 | 200 | 29.2 | 88 | 199 | 28.9 | 86.69 |
| 5 × 8 inter + intra | 240 | 201 | 28.6 | 90 | 199 | 28.04 | 91.83 |

* denotes variable logic depths.

## 2.5. Estimation of the design space

Using the proposed models, we can estimate the design space for the mean and standard deviation of different stages of a pipeline that will satisfy a yield constraint. Assuming the overall delay of the pipeline to be Gaussian and using (3), the upper bound of the mean of each stage delay ($\mu_i$) is given by:

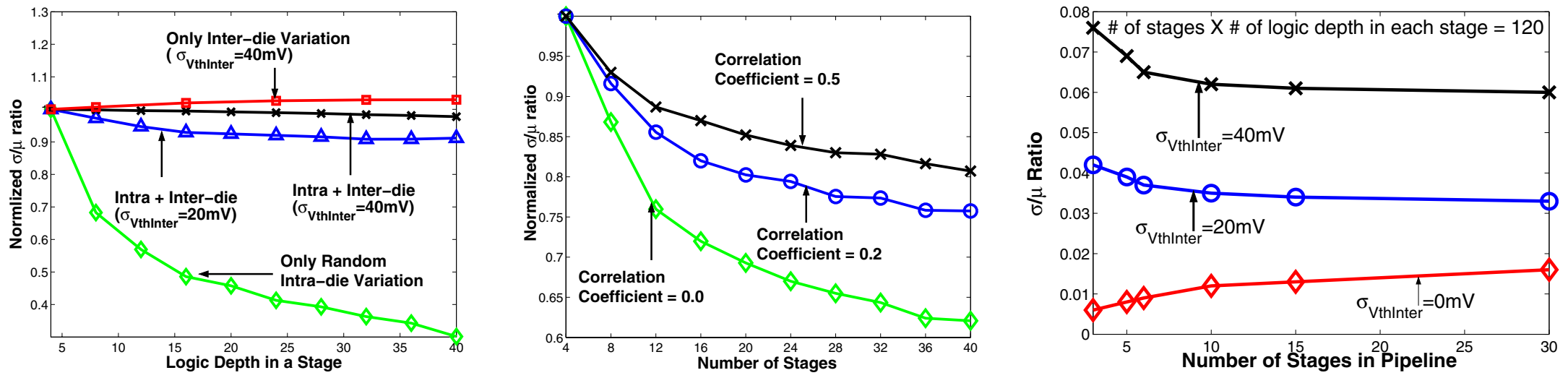

(a) (b) (c)

**Figure 5: Variability of (a) stage delay with logic depth, (b) pipeline design delay with the number of stages, (c) pipeline design delay with the simultaneous change of logic depths and number of stages (number of stages × stage logic depth = 120)**

$$\mu_i \le \mu_T \le T_{TARGET} - \sigma_T \Phi^{-1}\left(P_D\right) \quad (10)$$

However, this bound does not provide any estimate of the design space of $\sigma_i$ of each stage. A relaxed upper bound on the $\mu_i$ and $\sigma_i$ for the i[th] stage can be obtained (Fig. 4) by assuming that any stage j≠i meets the yield requirement with probability 1 (since maximum value of Φ(x) =1) as shown below:

$$\mu_i + \sigma_i \Phi^{-1}\left(P_D\right) \le T_{TARGET} \quad (11)$$

Equation (11) states that, if the mean and/or the standard deviation of any stage falls outside this bound, no pipelined design with that stage can ever meet the target delay and yield. A more stringent bound is obtained by assuming uncorrelated and equal stage delays and is given by:

$$\left[\Phi\left(\frac{T_{TARGET} - \mu_i}{\sigma_i}\right)\right]^{N_S} \ge P_D \Rightarrow \mu_i + \sigma_i \Phi^{-1}\left(P_D^{1/N_S}\right) \le T_{TARGET} \quad (12)$$

where, $N_S$ is the number of pipeline stages. Depending on $N_s$, it gives the set of values for $\mu_i$ and $\sigma_i$ that can meet the target yield (Fig. 4 shows 2 such equality bounds for $N_s = n1$, $n2$ with $n1 < n2$). Note that, there is a minimum bound on the $\mu_i$ and $\sigma_i$ which depends on the minimum allowable logic depth and process specification (Fig. 4). Moreover, the $\mu_i$ and $\sigma_i$ of a combinational circuit are related parameters and the relation determines the realizable design space for the $\mu_i$ and $\sigma_i$. For example, if we model each stage delay to be a chain of $N_L$ inverters then a simple relation between $\mu_i$ and $\sigma_i$ is given by:

$$\mu_i = N_L \mu_{\min} \text{ and } \sigma_i = \sqrt{N_L}\sigma_{\min} \Rightarrow \mu_i = \left(\mu_{\min}/\sigma_{\min}^2\right)\sigma_i^2 \quad (13)$$

where, $\mu_{min}$ and $\sigma_{min}$ are the mean and the standard deviations of an minimum sized inverter, respectively. Similarly for maximum sized inverter having parameters $\mu_{max}$ and $\sigma_{max}$, there will be another bound on realizable $\mu_i$ and $\sigma_i$ (Fig. 4, curve marked Realizable upper bound). Hence, for equal critical stage delays a realizable design region bounded by a set of curves given by equation (12) to (13) is obtained (Fig. 4).

# 3. Observations on statistical pipeline design

Using the analytical models presented in section 2, we have analyzed the effect of logic depth and the unbalancing of the stage delays on the pipeline yield.

## 3.1. Trade-off between number of pipeline stages and logic depth

The delay of a pipelined design depends on the logic depth of the pipe stages. A reduction in the logic depth, which increases the number of stages in a pipeline, improves the operating frequency [4]. In this section we analyze the effect of logic depth and number of stages on the variability of a pipeline delay. A design with a lower variability has a higher probability of meeting a target delay (i.e. better yield) [5].

If only random intra-die variation is considered, the delays of the different gates in a combinational logic behave as independent variables. Under this condition, increasing logic depth reduces the variability ($\sigma/\mu$) due to cancellation effect (i.e. delays of some gates increase while others decrease resulting in lower overall change) (Fig. 5(a)) [5]. However, correlation in the delays of the different gates (due to inter-die variation and spatial correlation) minimizes the cancellation effect. Hence, the variability becomes a weaker function of the logic depth (Fig. 5(a)). On the other hand, increasing the number of elements in the *max* function reduces its variability (Fig. 5(b), evaluated using a inverter chain pipeline with constant logic depth in each stage [7]. It can be further, observed that as the *stage delays* becomes more and more correlated, the sensitivity of the variability to the number of stages reduces (Fig. 5(b)).

In order to understand the effect of logic depth ($N_L$), number of stages ($N_S$) and relative strength of inter-die and intra-die variations, we have estimated the variability of an inverter chain pipeline (in BPTM 70nm technology node) of 120 levels with different configurations (i.e. with different $N_L$ and $N_S$ such that $N_L$ x $N_S$ =120). In each case the delay distribution proposed by the analytical models closely match the Monte-Carlo simulation results from SPICE. With only intra-die variation, the effect of logic depth prevails over the effect of the max function. Consequently, increasing the number of stages (i.e. reducing the logic depth of each stage) increase the variability (Fig. 5(c)). On the other hand, as the strength of inter-die variation increases (i.e. stage delays become more correlated) the σ/μ ratio of each stage becomes a weaker function of its logic depth. Hence, the impact of *max* function prevails and variability of the overall pipeline delay reduces with an increase in the number of stages (Fig. 5(c)).

## 3.2. Perfectly balanced vs. unbalanced pipeline design

Traditionally, the pipeline stages are designed for equal delay to maximize the throughput [4]. However, incorporating imbalance among the pipeline stage delays can have a positive impact on the yield under process variation. This is because of the fact that a balanced pipeline has more number of critical paths than an unbalanced design that adversely affects the yield [5]. Imbalance can be incorporated in a balanced pipeline by using transistor sizing and/or logic re-structuring. However, the impact of imbalance on the overall pipeline delay needs to be estimated.

We have performed experiments with a 3-stage pipeline structure to understand the effects of imbalance on the pipeline delay. For example, Fig. 6 shows a three-stage pipelined ALU-Decoder circuit. The combinational logic of each stage is first optimized for minimum area (using [3]) for a specific target pipeline delay (179ps) and yield (80%). First, we kept the target delay and the target yield constant for all the three stages (i.e. neglecting any correlation, the yield target for each stage was kept at $(0.80)^{1/3}$ = 0.9283 using (12)). In the next step, we have introduced imbalance among the three stages (by transistor sizing) in such a way that the total area remains constant. The overall delay distributions and the yield of the balanced and unbalanced pipeline are shown in Fig. 7.

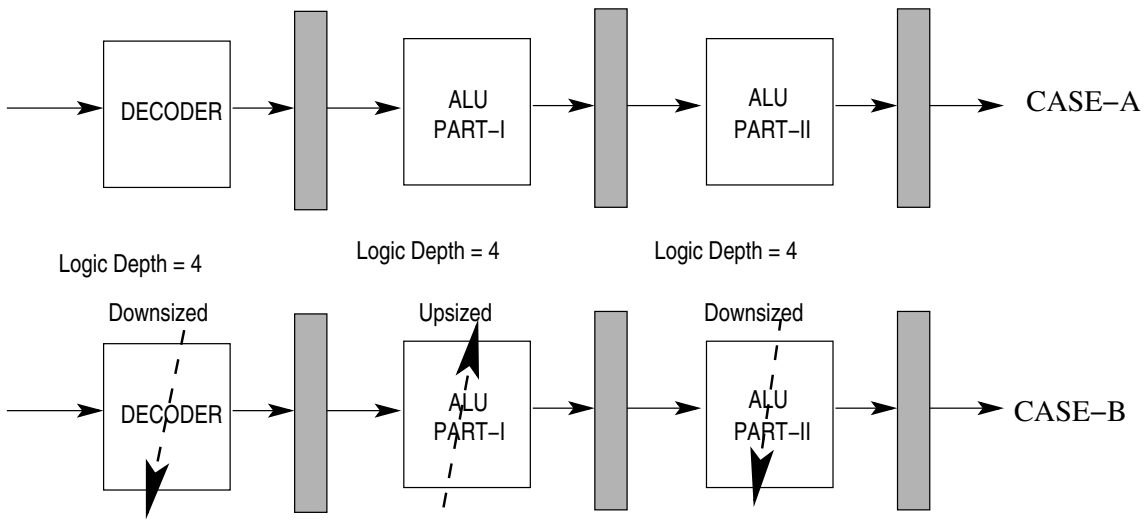

**Figure 6: A 3-stage pipeline with different stage delays**

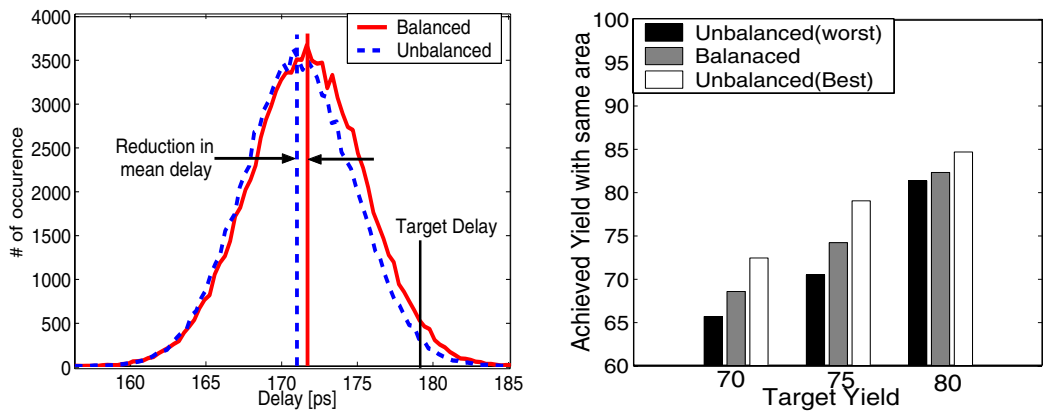

(a) (b)

**Figure 7: Effect of unbalancing on a) pipeline delay, b) yield**

Resizing the transistors reduces the mean but increases variance in pipeline delay (Fig. 7(a)). To understand the reason behind this yield improvement, let us consider the area vs. delay curves for each stage (Fig. 8). They are initially designed for equal delays indicated by line L1 in Fig. 8. This results in yield of $Y_0$ for each stage (pipeline yield = $Y_0^3$). The total area for this design is the sum of the stage areas (A1+A2+A3). Now, we introduce imbalance by reducing the area of stage 1 and 3 (by dA1 and dA3) increasing their delays to line L2. This reduces yields of stages 1 and 3, to $Y_1$, $Y_3$ (both less than $Y_0$). However, this extra area (dA1+dA3) can be added to the stage 2, thereby reducing its delay to line L3. This improves the yield of stage 2 (i.e. $Y_2 > Y_0$). If, in this particular case, $(Y_1 \times Y_2 \times Y_3) > Y_0^3$ the overall pipeline yield improves. This trend has been observed in the 3 stage pipeline circuits as shown in Fig. 7(b). However, introducing excess imbalance in stage delays, we might get diminishing returns when pipeline performance is governed by the $\mu$ of the slowest stage (Fig. 7(b), worst case unbalancing results).

Hence, it is necessary to appropriately apply imbalance among the stages. Using the above observations, we have developed a simple heuristic to provide imbalance among the stages as shown below:

*Calculate rate of change of area with delay for each stage* : $R_i = \left|\frac{\partial A}{\partial D}\right|_i$

*For each stage*
  *if* $R_i > 1$
    *reduction in large area*
    *results in small increase in delay.* (14)
  *else*
    *increase in small area*
    *results in large improvement in delay.*
  *endif*
*endfor*

In order to improve the yield of a design with minimum impact on area, the delays of the stages with ($R_i < 1$) should be more effective to reduce (small area penalty). On the contrary, to reduce area (or power) with minimum penalty on yield, the area of the stages with ($R_i > 1$) should more effective to reduce (small delay increase).

# 4. Pipeline design flow under statistical delay distribution

Using the models developed in section 2 and based on the observations in section 3, we propose a statistical design method for the complete pipelined circuit. The proposed method is targeted to optimize area (hence, power) while ensuring the yield. In a conventional pipeline design flow, individual stages are designed and optimized independently of others before they are glued together to form a pipeline.

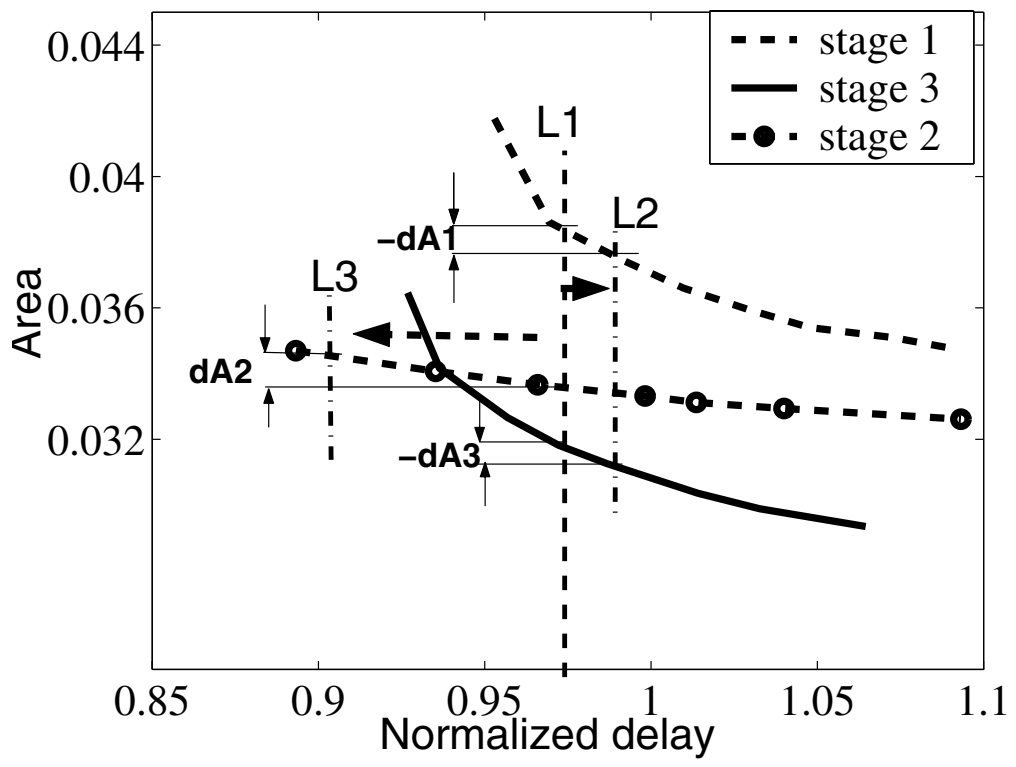

**Figure 8: Area vs. delay curves of different logic stages of the 3-stage ALU-Decoder pipeline design**

Under statistical delay variation, a global optimization of the complete pipeline is necessary for two reasons:

- Yield target ($Y$) for the complete pipeline may not be met if some of the stages fail to meet their yield target ($Y_0$). In that case, the delay of the stages meeting yield target $Y_0$ can be further improved to compensate for the stages failed to meet the yield.
- Although the yield constraint is satisfied for each stage, the integrated pipelined circuit has opportunity for optimization with respect to the area/power (section 3.2).

## 4.1. Global optimization of the pipeline for target yield

The optimization problem for minimizing area of a pipeline under a given yield constraint can be formulated as below,

Minimize

$\sum_{i=1}^{N} Area\,(SD_i)$ /* $SD_i$ stands for delay of the $i$-th stage */

Subject to

$\phi\left(\frac{T_{TARGET} - \mu_T}{\sigma_T}\right) \geq Y$ /* $Y$ is target yield */

$L_i \leq x_i \leq U_i \quad i = 1,....,n.$ /* $x_i$ is the size factor for a logic gate. $L_i$ and $U_i$ are the minimum and maximum size factors */

To solve the above problem, we use gate-level sizing algorithm for minimizing total area under yield constraint given in [3]. It is an iterative low-complexity algorithm based on Lagrangian Relaxation (LR) [3]. We have developed a global design optimization algorithm to solve the above problem efficiently (Fig. 9). Application of the proposed algorithm directly to the complete pipeline, where all the stages are sized simultaneously, is computationally very expensive. This is because of the fact that, a pipelined circuit can be very large and will take inordinately large run time and storage to converge [3]. The algorithm employs the principle of divide-and-conquer where we size one stage at a time in such a way that the target yield for the complete pipeline is satisfied while the total area is minimized. It should be noted that, for a pipelined design with a target yield $Y$ with respect to a delay of $T_{TARGET}$, the yield constraint to each stage is now set to $Y$ (for $T_{TARGET}$). Moreover, statistical delay analysis is performed over the complete pipeline, although the sizing is done for only one stage. It helps to make the algorithm computationally efficient, since we avoid application of the sizing routine on all the stages simultaneously.

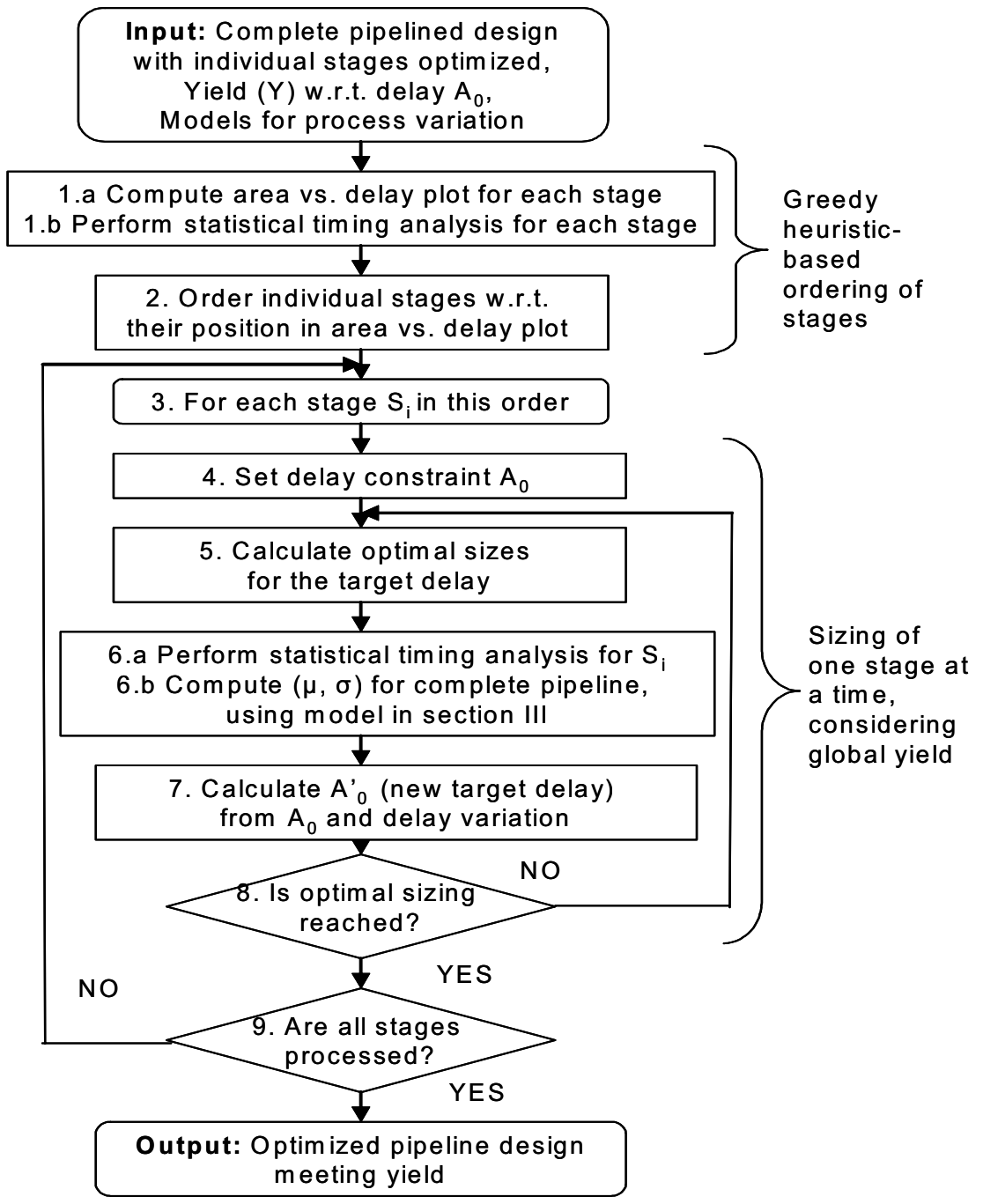

**Figure 9: Algorithm for the global optimization of pipeline for ensuring yield with minimum area overhead**

It is worth noting that, improvement in the total area of the pipeline strongly depends on the ordering at which the stages are chosen for sizing. In our algorithm, the ordering is based on the position of each stage in their area vs. delay curve as explained in section 3.2 (14). This minimizes the total design area to meet the target pipeline yield during global optimization, since stages with lower $R_i$ are sized before stages with higher $R_i$.

For each stage, we apply the algorithm in [3], which starts with assigning an initial delay constraint $A_o$ (pipeline delay) to the stage and iteratively find optimal sizes with respect to it. At the end of each iteration, statistical timing analysis on the complete pipeline is performed. Depending on the new $\mu$ and $\sigma$ of the pipeline delay, the pipeline target delay is modified to $A'_o$ (steps 4-7 in Fig. 9). Since logic gates of only one stage are being sized at a time, we perform full timing analysis on one stage only to estimate its mean and the standard deviation. The proposed analytical models (section 2.2) are then used to produce the resulting pipeline delay distribution. This incremental timing analysis further improves the computational efficiency of the algorithm.

**Table-II Ensuring $Y_{TARGET}$ (80%) with small area penalty**

| Stage Logic | Individually Optimized | | Proposed Algorithm | |
|---|---|---|---|---|
| | Area (%) | Yield (%) | Area (%) | Yield (%) |
| c3540 | 47.4 | 86.3 | 47.3 | 86 |
| c2670 | 25.7 | 95 | 27.4 | 99.1 |
| c1980 | 20.4 | 95 | 20.7 | 95.6 |
| c432 | 6.5 | 95 | 6.6 | 99.2 |
| Pipeline: | 100 | 73.9 | 102 | 80.5 |

**Table-III Area reduction for a target yield (80%)**

| Stage Logic | Individually Optimized | | Proposed Algorithm | |
|---|---|---|---|---|
| | Area (%) | Yield (%) | Area (%) | Yield (%) |
| c3540 | 50 | 94 | 45 | 90.5 |
| c2670 | 23.2 | 95 | 21.2 | 99.1 |
| c1980 | 20.3 | 95 | 19.1 | 90.5 |
| c432 | 6.5 | 94.5 | 6.3 | 99.2 |
| Pipeline: | 100 | 80.3 | 91.6 | 80.5 |

The global optimization algorithm proposed here is significantly faster and takes much less storage compared to the case where all the stages are sized simultaneously. The LR based sizing algorithm proposed in [3] has a computational complexity of $O(n^2)$ where $n$ is the number of logic gates to size. For $m$ pipeline stages each having $n$ gates the simultaneous sizing approach runs with a complexity of $O(m^2n^2)$ (with space complexity of $O(mn)$). The proposed algorithm improves the complexity to $O(mn^2)$ (with space complexity of $O(n)$).

The algorithm is applied on an example 4-stage pipelined circuit (stages are designed with ISCAS85 benchmark circuits) and the results were compared with the case where the logic stages were individually optimized using method in [3]. We observed that, preferentially incorporating imbalance among the stages using the proposed algorithm ensures target yield (Table-II) or minimizes area for a target yield (Table-III). The highlighted rows in the tables show the stages chosen by the algorithm for yield improvement with very small area increase. Similarly, the plain rows (Table-II and III) show stages chosen by the algorithm for area saving with small penalty in yield. Overall a 9% improvement in yield can be obtained with a small area penalty 2% (Table-II). On the other hand, about 8.4% area improvement can be obtained for the same yield (table-III).

## 5. Conclusions

We have investigated pipeline delay distribution under inter- and intra-die parameter variations. Analytical models for estimating yield of a pipelined circuit are presented. We have observed the impact of logic depth (or number of pipe stages) and imbalance among the stage delays on the variability of the pipeline delay. An efficient sizing algorithm for pipeline to minimize area under yield constraint is presented. Our investigations show that a statistical design of a complete pipeline (not only the individual stages) is effective to improve yield in presence of parameter variations.

## REFERENCES

[1] K. A. Bowman et al., "Impact of Die-to-Die and Within-Die Parameter Fluctuations on the Maximum Clock Frequency Distribution for Gigascale Integration", *JSSC* 2002, pp. 183-190.

[2] E. Jacobs et al., "Gate Sizing Using a Statistical Delay Model", *DATE 2000,* pp. 283-290.

[3] S. Choi, et al., "Novel Sizing Algorithm for Yield Improvement under Process Variation in Nanometer Technology", *DAC 2004*, pp. 454-459.

[4] J. L. Hennessy et al., "Computer Architecture: A Quantitative Approach", *Morgan Kaufmann*, 3-rd edition, May 2002.

[5] S. Borkar et al., "Parameter Variations and Impact on Circuits and Microarchitecture", *DAC 2003*, pp. 338-342.

[6] H. Mahmoodi, et al., "Estimation of Delay Variations Due to Random-dopant Fluctuations in Nano-Scaled CMOS circuits", *CICC 2004,* pp. 17-20.

[7] A. M. Ross, "Useful Bounds on the Expected Maximum of Correlated Normal Variables", Aug 2003.

[8] C. E. Clark, "The Greatest of a Finite Set of Random Variables", *Operations Research 9(2)*, Mar-Apr, 1961, pp. 145-162.

[9] BPTM: http://www-device.eecs.berkeley.edu/~ptm/